\documentclass{IEEEtran}
\usepackage{amssymb}
\usepackage{amsfonts}
\usepackage{amsmath}
\usepackage{cite,multicol,multirow,diagbox,booktabs,array}
\usepackage{graphicx}
\usepackage{subfig}
\usepackage{lipsum}
\usepackage{verbatim}
\usepackage{algorithmic}
\usepackage[justification=raggedright]{caption}
\usepackage {pifont}
\usepackage[ruled,vlined]{algorithm2e}
\usepackage{enumerate}

\usepackage[dvips]{color}
\usepackage{float}

\newtheorem{remark}{Remark}

\newtheorem{assumption}{Assumption}
\newtheorem{definition}{Definition}

\usepackage{etoolbox}
\pretocmd{\maketitle}{%
  \markboth{The definitive version was published in IEEE Wireless Communications Letters, vol. 11, no. 7, pp. 1364-1368, July 2022, doi: 10.1109/LWC.2022.3169215.}{}%
}{}{}

\ifCLASSOPTIONcompsoc
\fi
\ifCLASSINFOpdf
\else
\fi
\begin{document}
\title{Federated Learning-Based  Localization  with Heterogeneous Fingerprint Database}
\author{Xin Cheng,~Chuan Ma,~Jun Li,~\IEEEmembership{Senior member,~IEEE},~Haiwei Song,~Feng Shu,\\ and~Jiangzhou Wang,~\IEEEmembership{Fellow,~IEEE}
\thanks{
Xin Cheng, Chuan Ma, and Jun Li are with School of Electronic and
Optical Engineering, Nanjing University of Science and Technology, Nanjing,
210094, China. (e-mail:xincstar23@163.com).

Haiwei Song is with the 8511 Research Institute, China Aerospace Science and Industry Corporation, Nanjing, 210007, China.

Feng Shu is with the School of Information and Communication Engineering,
Hainan University, Haikou, 570228, China and with the School
of Electronic and Optical Engineering, Nanjing University of Science and
Technology, Nanjing, 210094, China.

Jiangzhou Wang is with the School of Engineering, University
of Kent, Canterbury CT2 7NT, U.K.}}
\maketitle
\begin{abstract}
  Fingerprint-based localization plays an important  role in indoor location-based services, where the position information is usually collected in distributed clients and gathered in a centralized server. However, the overloaded transmission as well as the potential risk of divulging private information burdens the application.
  Owning the ability to address these challenges,  federated learning (FL)-based fingerprinting localization comes into people's sights, which aims to train a global  model while keeping raw data locally. However, in distributed machine learning (ML) scenarios, the unavoidable database heterogeneity usually degrades the performance of  existing FL-based localization algorithm (FedLoc). In this paper, we first characterize the database heterogeneity with a computable metric, i.e., the area of convex hull, and verify it by  experimental results.  Then, a novel heterogeneous FL-based localization algorithm with the area of convex hull-based aggregation (FedLoc-AC) is proposed.  Extensive experimental results, including real-word cases are conducted. We can conclude that the  proposed FedLoc-AC can achieve an obvious prediction  gain compared to FedLoc in  heterogeneous scenarios and has  almost the same prediction error with it  in   homogeneous scenarios. Moreover, the extension of FedLoc-AC in multi-floor cases is proposed and verified.
\end{abstract}


\IEEEpeerreviewmaketitle

\section{Introduction}
The explosion of smart devices and the ever-growing sensing and computing technologies have motivated  the development of indoor location-based services (LBS)\cite{8409950}. With the coming of beyond 5G (B5G) and internet of things (IoT), LBS becomes indispensable.  However, it is   challenging to achieve high localization accuracy using traditional localization approaches. For example, the weak signals emitted from satellite cannot work well in indoor environments,  making global navigation satellite system  unserviceable.   Moreover, other empirical and model-based technologies  mismatch the underlying mechanism of complex indoor environments. Owning the ability of remedying these defects, the technique of received signal strength (RSS) fingerprint-based indoor positioning has received increasing attentions\cite{7174948}.

An attractive solution for such localization is the centralized machine learning (ML) algorithm\cite{8726079}.  In the off-line phase, a site survey is performed with clients by measuring the strength pattern of signals at different sampling positions in the area of interest (AoI). Such  signals are emitted from access points (APs), including WiFi, Bluetooth,  Zigbee, etc. Then, a ML model is trained in the server to map the RSS vector and the corresponding measuring position. In the on-line phase, a user  can query its position by inputting the real-time measured signal pattern to the trained model.

However,  in the off-line phase,   clients are required to  send the raw data to the server for model training, causing   disclosure of clients's  position information. The privacy protection issue severely hinders the promotion and scalability of LBS\cite{8726079}.  Besides, with a large-scale raw data and  client amount,  the gathering process leads to a high communication cost.

To alleviate the over-loaded communication cost and protect the client privacy,  federated  learning (FL) has been introduced to the fingerprint-based localization system\cite{9148111,9250516,9103044}. The gist of FL is to learn a global model in a distributed manner  while keeping raw data locally, and  only  model parameters are exchanged between clients and server. Therefore, FL-based localization is a promising approach to address the above-mentioned challenges. The existing FL-based localization algorithms \cite{9148111,9250516,9103044} are based on Federated Averaging\cite{A17}. However, it shows that the heterogeneous  nature of fingerprint database limits the prediction performance of these algorithms.

In detail, in practical fingerprint-based localization applications, database heterogeneity usually emerges due to the unbalanced client  behaviour. For example,
\begin{enumerate}
  \item Smart devices may have different hardwares, e.g., battery, sensor and computing  unit.
  \item Different environmental factors like obstacles may differ  sample states of  smart devices.
  \item Smart devices may be ordered/taken by different owners, causing various moving states.
\end{enumerate}
Such unbalanced behaviours will result in unbalanced sampling characteristics in the AoI, e.g.,  sampling intervals, sampling amount and trajectories. Consequently, the database heterogeneity is generated, which is mainly reflected by the spatial distribution of sampling positions in the AoI. Therefore, the heterogeneous characteristic of fingerprint database should not be omitted in the design of localization algorithms, which has been neglected in existing FL-based localization algorithms\cite{9148111,9250516,9103044}.
In this paper, we first characterize the heterogeneity of the fingerprint database, and then a novel  heterogeneous FL-based localization algorithm with the area of convex hull-based aggregation (FedLoc-AC) is proposed.  To the best of our knowledge, this is the first work to characterize the heterogeneity of fingerprint database and design FL-based localization algorithm with it. The main contributions of this paper are summarized as follows:
\begin{enumerate}
\item The fingerprint database heterogeneity  exists in distributed localization tasks.  To characterize it, we associate  the heterogeneity with the spatial distribution of sampling positions, and  propose a computable heterogeneous characteristic, i.e.,  the convex hull of  sampling positions.
\item To improve the localization accuracy,  a novel FL-based localization algorithm, named as FedLoc-AC, is proposed by elaborating the proposed heterogeneous characteristic of fingerprint database.  Besides, the convergence property of the FedLoc-AC is provided. Moreover, the adaption of FedLoc-AC in multi-floor cases is proposed.
\item   We conduct extensive experiments to verify the effectiveness of the proposed heterogeneous characteristic.
   In addition, experimental results, including real-word and 3D cases show the  proposed FedLoc-AC can achieve considerable prediction  gain compared to FedLoc in  heterogeneous scenarios and has almost the same prediction error with it in   homogeneous scenarios.
\end{enumerate}

The rest of this  paper is structured as follows. In Section II, we describe the process of  FL-based localization and address the problem of fingerprint database heterogeneity. In Section III, the  fingerprint database heterogeneity is characterized, and then the FedLoc-AC is proposed based on it.  We present experimental results in  Section IV ,and   conclude this paper in Section V.

\section {FL-based Localization}
\subsection{Fingerprint Database Construction}
Consider a  fingerprint-based indoor localization  with the assistance of $N$  clients and  $L$ APs in the AoI.   These APs are deployed at fixed positions  to broadcast WiFi beacons. The clients here refer to  smart devices equipped with sensing, logging, storage, computing and communication entities.
To perform the site survey in the AoI, each client  moves to a certain number of  positions to read the strength pattern of received signals emitted from APs. The RSSs sampled by  the $i$-th client at a position can be formulated in a vector form as
\begin{align}
\mathbf{x}_{i}=\begin{bmatrix} \mathrm{RSS}_{i,1},&  \mathrm{RSS}_{i,2}&   ,\cdots, &\mathrm{RSS}_{i,L} \end{bmatrix},
\end{align}
where $\mathrm{RSS}_{i,L}$ denotes the measured RSS  at the $i$-th client,  from the $j$-th AP.

After finishing the site survey, the $i$-th client storages $M_{i}$ data pairs, consisting of  measured RSSs and corresponding positions. Subsequently, a local database, denoted as $\mathcal{D}_{i}$ is constructed at the $i$-th client, which is expressed as
\begin{align}\label{dataset}
\mathcal{D}_{i}=\{(\mathbf{x}_{i}^{1},\mathbf{y}_{i}^{1}), (\mathbf{x}_{i}^{2},\mathbf{y}_{i}^{2}), \cdots, (\mathbf{x}_{i}^{M_{i}},\mathbf{y}_{i}^{M_{i}})\},
\end{align}
where $\mathbf{y}_{i}$ is the corresponding measuring position of  $\mathbf{x}_{i}$.

\subsection{Federated Learning Process}
\begin{figure}
  \setlength{\belowcaptionskip}{-0.4cm}
  \centering
  \includegraphics[width=0.45\textwidth]{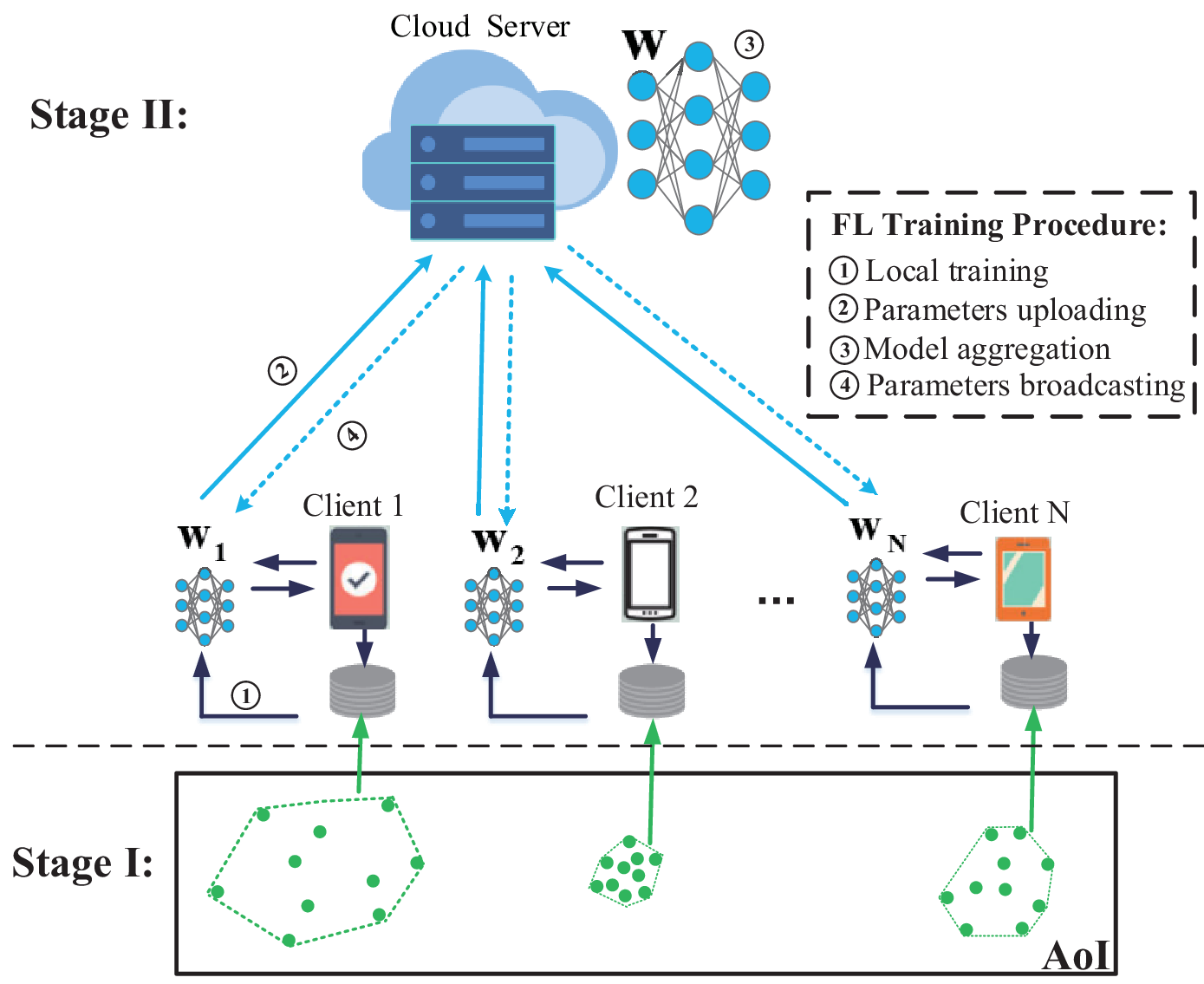}\\
  \caption{Block diagram of FL-based localization.
  Stage I: Fingerprint database construction (the green points represent the sampling positions in the site survey and the dotted line is the edge of corresponding convex hull).
  Stage II: The FL work flow. }\label{flow}
\end{figure}
In this subsection, we present the process of training a  multiple perceptron (MLP) model for localization in a federated manner.

For ease of understanding, a complete picture of FL framework is depicted in Fig.~\ref{flow}.
The FL process consists of  information interactions between the central server and clients with $T$ global epoches. At the $t+1$-th round, the process of such a FL system contains the following four steps:
\begin{itemize}
  \item \textit{Local training:} each client completes the local update on the  global model  $\mathbf{w}^{t}$, received from the central server. The update is based on  optimizing the model over the local fingerprint dataset. The  mean absolute error (MAE) is selected  as the local loss function,  defined as
  \begin{subequations}\label{localobj}
  \begin{equation}
  \mathcal{L}_{i}(\mathbf{w})=\frac{1}{M_{i}}\sum_{m=1}^{M_{i}}l_{i}(\mathbf{x}_{i}^{m},\mathbf{y}_{i}^{m};\mathbf{w}),
  \end{equation}
  \begin{equation}
  l_{i}(\mathbf{x}_{i}^{m},\mathbf{y}_{i}^{m};\mathbf{w})=\|\mathcal{F}(\mathbf{x}_{i}^{m};\mathbf{w})-\mathbf{y}_{i}^{m}\|_{2},
  \end{equation}
  \end{subequations}
  where   $\mathbf{w}$ is the model to be optimized and  $\mathcal{F}$ is the model function. The local update  is finished by $E$ steps of stochastic gradient descent at $\mathbf{w}^{t}$ with a learning rate $\eta$.  Let  $\mathbf{w}_{i}^{t+1}$ denote the local updated model of the $i$-th client.

  \item \textit{Parameters uploading:} all  clients transmit the local updated model   to the central server.

  \item \textit{Model aggregation:} the central server aggregates the uploaded local model and generates new global model, expressed as
\begin{align}
\mathbf{w}_{i}^{t+1}=\sum_{i=1}^{N} p_{i}\mathbf{w}_{i}^{t+1},
\end{align}
where $p_{i}$ represents the aggregating weight of the $i$-th client with $\sum_{i=1}^{N} p_{i}=1$.

  \item \textit{Parameter broadcasting:} the central server broadcasts the aggregated $\mathbf{w}^{t+1}$ to all clients for the next round of learning.

\end{itemize}
\subsection{Fingerprint Database Heterogeneity}
The  objective function in the FL-based localization can be formulated as
\begin{align}\label{obj}
\mathcal{L}(\mathbf{w})\triangleq \sum_{i=1}^{N} p_{i}\mathcal{L}_{i}(\mathbf{w}).
\end{align}
It implies that $p_{i}$ plays an important role in the prediction performance of  trained model.

In practical  applications, database heterogeneity usually emerges due to unavoidable unbalanced device behaviours. Therefore the FL-based localization should be designed by considering this vital factor. Furthermore, the aggregating weights should be determined by the heterogeneity characteristic of fingerprint databases. As illustrated in Fig.~\ref{flow}, the spatial distribution of sampling positions is an important heterogeneity characteristic. For example, even with the same data size, the coverage area of fingerprint database differs from each other. In the next section, we will characterize the spatial distribution of fingerprint database by  a well-known definition of convex hull\cite{boyd2004convex}. Then we will elaborate this characteristic to the aggregating design of FL-based localization.


\section{The Proposed FedLoc-AC}
\subsection{Heterogeneous Characteristic of Fingerprint Database}
\begin{figure}
  \setlength{\belowcaptionskip}{-0.5cm}
  \centering
  \includegraphics[width=0.45\textwidth]{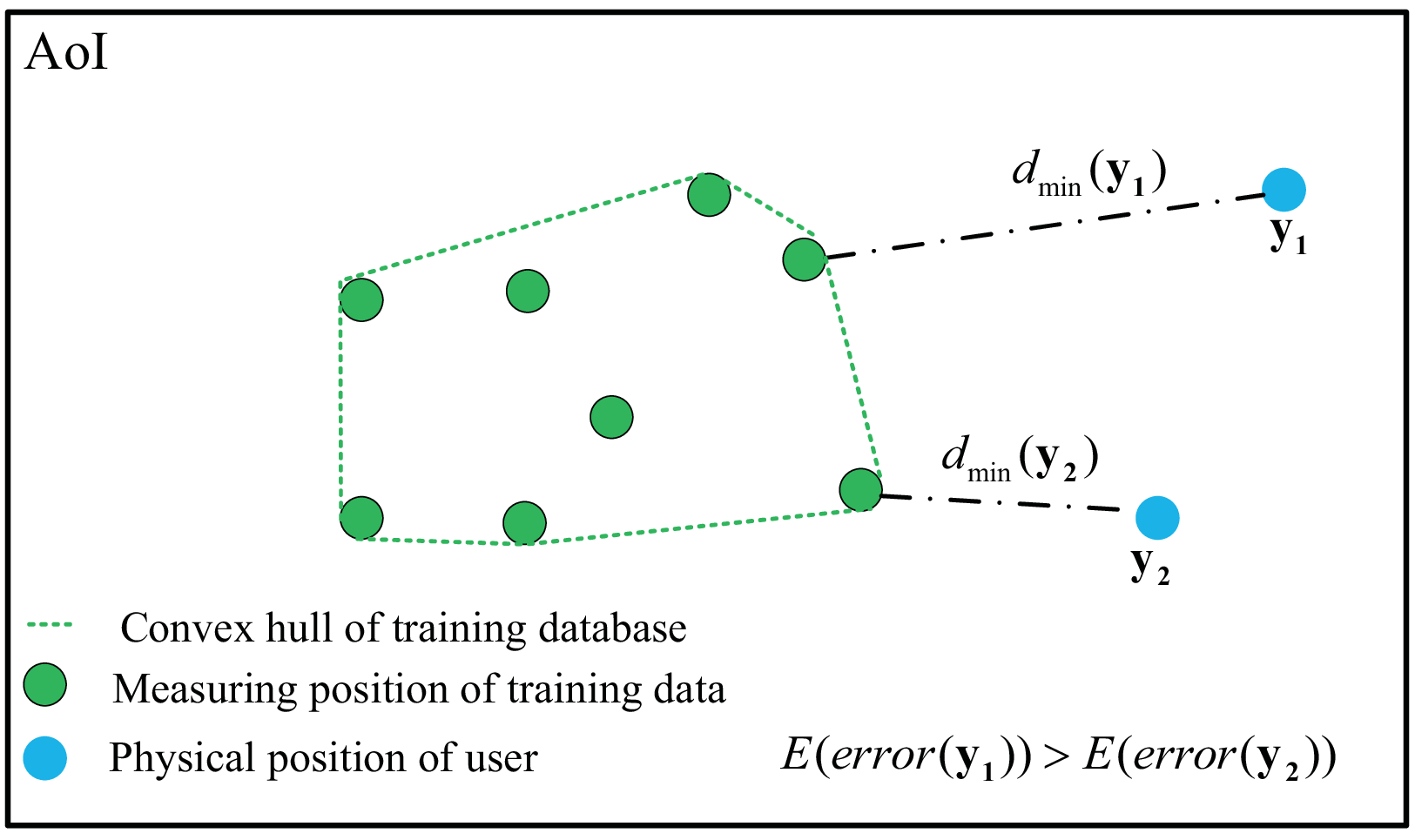}\\
  \caption{The relationship between the spatial distribution of  fingerprint database and the prediction error.}\label{spatial}
\end{figure}

In Fig.~\ref{spatial}, we further explain the  heterogeneous characteristic of fingerprint database, i.e., the spatial distribution of sampling positions.   In detail, in the online phase, a user at position $\mathbf{y}$ is querying its position by inputting its real-time measured RSS vector to a MLP model, which is trained  by the fingerprint database $\mathcal{D}$ in the off-line phase. Let  $\mathbf{y}^{'}$ denote the single measuring position in  $\mathcal{D}$  and $\mathcal{D}(\mathbf{y}^{'})$ denote the set of whole sampling positions in $\mathcal{D}$. Without loss of generality, the average prediction error at position $\mathbf{y}$, denoted as $e(\mathbf{y})$, is proportional  to the minimum distance between $\mathbf{y}$ and $\mathcal{D}(\mathbf{y}^{'})$, expressed as
\begin{align}\label{mderror}
e(\mathbf{y})\propto  d_{min}(\mathbf{y};\mathcal{D}(\mathbf{y}^{'}))=\min_{\mathbf{y}^{'}\in \mathcal{D}(\mathbf{y}^{'})}{\|\mathbf{y}-\mathbf{y}^{'}\|_2}.
\end{align}

After illustrating the effect of the spatial distribution to the estimation of a single position, we will expand it to the whole AOI. Since the  mainly concerned performance of the trained model is the average prediction accuracy over the AoI, a comprehensive  performance metric is given by
\begin{align}\label{AoIe}
\mathcal{E}=\int_{\mathbf{y}\in AoI}  e(\mathbf{y}) f(\mathbf{y}) d \mathbf{y}=\frac{1}{S} \int_{\mathbf{y}\in AoI}  e(\mathbf{y}) d \mathbf{y},
\end{align}
where  $f(\mathbf{y})$ is the probability density function of  user position in the AoI and $S$ is the area of the AoI. We consider a uniform distribution of user position here. Substituting (\ref{mderror}) into (\ref{AoIe}), we  obtain
\begin{align}
\mathcal{E}\propto \underbrace{\int_{\mathbf{y}\in AoI} \min_{\mathbf{y}^{'}\in \mathcal{D}(\mathbf{y}^{'})}{\|\mathbf{y}-\mathbf{y}^{'}\|_2} d \mathbf{y}}_{g(\mathcal{D}(\mathbf{y}^{'}))}.
\end{align}

Our goal is to find a performance metric of the fingerprint database $\mathcal{D}$  to evaluate its effect to $\mathcal{E}$. We introduce an experiential computable geometrical characteristic to  represent $g(\mathcal{D}(\mathbf{y}^{'}))$ with negative relationship. That is the area of the convex hull of sampling positions, which is represented by the smallest convex set that contains the sampling positions\cite{boyd2004convex}. Let $C$  denote the convex hull and $S_{C}$ denote the area of it. The convex hull of sampling positions of the $i$-th device  is given by
\begin{align}\label{convexhullf}
&C_{i} = \\ \nonumber
&\{\theta_{1}\mathbf{y}_{i}^{1}+\cdots+\theta_{M_{i}}\mathbf{y}_{i}^{M_{i}}, \theta_{k}\geq0, k=1,...,M_{i}, \sum_{k=1}^{M_{i}}\theta_{k}=1\}.
\end{align}
Note that the notations in (\ref{convexhullf}) follow the formula  (\ref{dataset}).

\subsection{FedLoc-AC}
In this subsection, we  propose a FedLoc-AC.  The gist of FedLoc-AC is to allocate the aggregated weights according to the area of convex hull of the fingerprint database, denoted as $S_{C_{i}},~i=1,2,\cdots, N$.  The aggregating rule for the $i$-th client is designed as
\begin{align}
p_{i}=\frac{S_{C_{i}}}{\sum_{i=1}^{N} S_{C_{i}}}.
\end{align}
By doing so, the fingerprint database with a larger overlay area will contribute more to the global training model. The proposed algorithm is outlined in  Algorithm.~\ref{EFedLocA}. Note that we simplify the client as $Cl$.

\begin{algorithm}
\caption{Heterogeneous FL-based localization}
\label{EFedLocA}
\LinesNumbered
\KwIn{$T$, $\mathbf{w}^{0}$, $\eta$, $E$, $\mathcal{D}_{i}~ \forall i$}
\KwOut{$\mathbf{w}^{T}$}

Preparation:

\While{$Cl_{i}\in {Cl_{1},Cl_{2},\ldots Cl_{N}}$}{
      Find out the convex hull of $\mathcal{D}_{i}$, denoted as $C_{i}$, using Melkman algorithm\cite{melkman1987line}

      Compute the area of $C_{i}$, i.e, $S_{C_{i}}$

      Send $C_{i}$ to the central server}

Initialization:  $t=1$.

\While{$t\leq T$}{ \
Local training process:

\While{$Cl_{i}\in {Cl_{1},Cl_{2},\ldots Cl_{N}}$}{\
$\mathbf{w} \leftarrow \mathbf{w}^{t-1}$

\ForEach{ $e \in {1,2\ldots E}$}{$\mathbf{w}\leftarrow \mathbf{w}-\eta \nabla \mathcal{L}_{i}(\mathbf{w})$\;}\
$\mathbf{w}_{i}^{t} \leftarrow \mathbf{w}$

Upload parameters $\mathbf{w}_{i}^{t}$}

Model aggregating process:

Update the global model parameters $\mathbf{w}^{t}$ as \
$\mathbf{w}^{t}=\sum_{i=1}^{N}\frac{S_{C_{i}}}{\sum_{i=1}^{N} S_{C_{i}}}\mathbf{w}_{i}^{t}$

The central server broadcasts global model parameters

$t=t+1$}
\end{algorithm}

\begin{remark}
The authors in  \cite{8664630} have derived the  convergence property of  federated learning  theoretically.   We find that the convergence property of FedLoc-AC  follows  \cite{8664630} by  some adjustments. More details can be found in the appendix.
\end{remark}

\begin{remark}
Compared with the FedLoc \cite{9250516} in computational complexity, the  FedLoc-AC  needs the $i$-th client to compute $S_{C_{i}}$  additionally,  whose complexity is $O(\mathcal{D}_{i})$.  Due to the co-existing  federated learning process in FedLoc and FedLoc-AC, such additional cost is negligible.
\end{remark}

\subsection{Extension to 3D Cases}
Recently, positioning in a 3D case, especially a multi-floor building has  attracted extensive  attentions. However, predicting the three dimensional position of a user in a multi-floor building  usually takes poor precision. An effective approach is to location by two stages. In the first stage, the floor of the user is located by a by a ML classifier. In the next stage,  the position of the user on the located floor is predicted by a ML model, functioning only for the located floor.

The FedLoc-AC can be adapted to this case under FL framework by only adding a floor classification step. Firstly, the floor classifier is trained  with  fingerprint databases of distributed clients in the multi-floor building by federated averaging algorithm (FedAvg)\cite{A17}. Then for each floor, a ML model is trained  with  fingerprint databases of distributed clients. The proposed method, e.g., FedLoc-AC, can be employed directly. With the floor classifier and several floor-specific ML  models, the position of user can be predicted, as illustrated the above paragraph.

\section{Experimental Results}
In this section, we evaluate the  performance of the proposed algorithm,  compared with the centralized MLP and FedLoc\cite{9250516} under different scenarios. Besides, the effectiveness of  proposed heterogeneous characteristic  is verified.

\textbf{Experimental Environment}: To construct typical and straightforward scenarios clearly, the simulated dataset are synthesized. The AoI is an indoor environment with $50\times 50~\mathrm{m}^2$.  There are four WiFi nodes fixed at the corner of the AoI, emitting electromagnetic wave with power $P_{0}$. The  propagation of electromagnetic wave in the  AoI is simulated according to the radio-channel propagation model in \cite{5759777}.  The related wireless propagation environment is set as follows. The transmit power of each AP is $10$ dBm, and the received power loss at the  $1$ m reference distance is  $-30$ dBm. To simulate the  complicated indoor environment, the path loss is set to range from 3 to 8 and the variance of noise  is set to range from 2 to 8 in the AoI.

We consider both  homogeneous and heterogeneous  scenarios. In heterogeneous scenario, we consider a typical unbalanced device behaviors, causing fingerprint database heterogeneity, i.e., moving velocity.  The details are showed as follows.

\textbf{Homogeneous scenario}: The data collection and model training are completed by $8$ clients. Starting from the  vertexes of the rectangle AoI, each client moves with a velocity of $0.5~\mathrm{m}/\mathrm{s}$ with a sampling interval of  $3$ s. At each sampling position, the sample is obtained after averaging over $10$ measurements.  After sampling at $200$ positions,  each client constructs the local fingerprint database, and then cooperates with a central server to train a global model. The test database is generated at randomly positions in the AoI for $1200$ times.

\textbf{Heterogeneous scenario}: The basic settings maintain the same with the homogeneous scenario, while a half of the participated clients act as straggles with a limited moving velocity of $0.05~\mathrm{m}/\mathrm{s}$, resulting in unbalanced sampling spacing.

\textbf{Learning  Structure}: TensorFlow libraries are utilized to implement the learning process by a MLP network.  The MLP is trained using  SGD consisting of a single hidden layer with $64$ hidden units, where ReLU units are selected as active function. We set $T=300$, $E=40$, and $ \eta=0.00001$.

\subsection{Verifications of the Heterogeneous Characteristic}
\begin{figure}[tbp]
    \setlength{\belowcaptionskip}{-0.5cm}
	\centering
	\subfloat[Average error via minimum distance]{\label{mde}\includegraphics[width=0.25\textwidth]{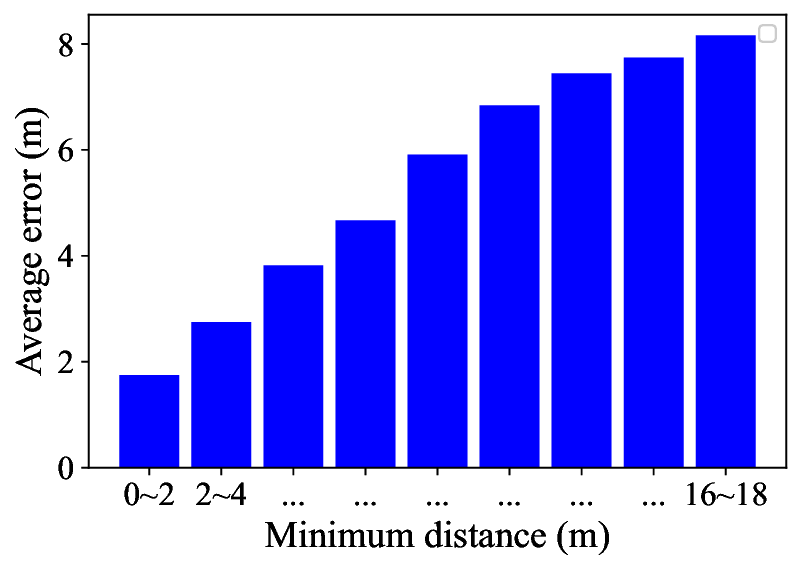}}
	\subfloat[Average MSE via area of convex hull]{\label{areae}\includegraphics[width=0.25\textwidth]{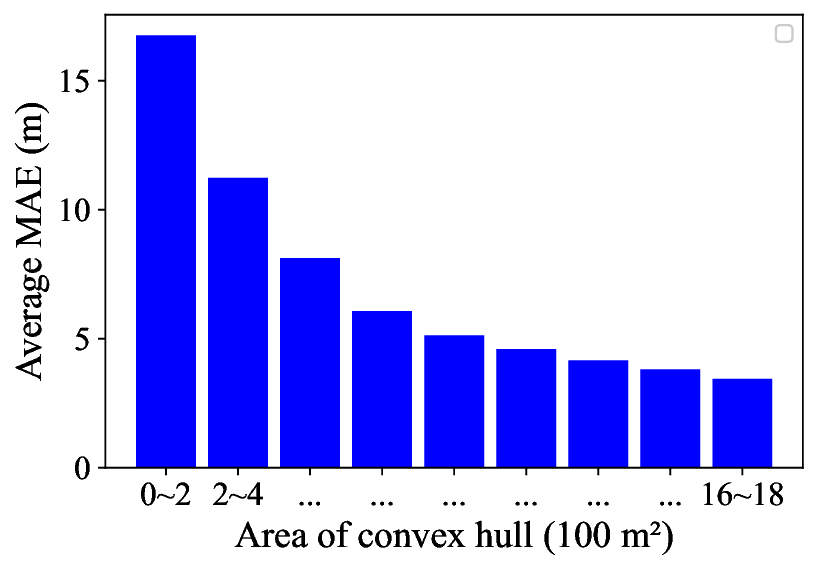}}
	\caption{Verifications of  the heterogeneous characteristic}
    \label{verification}
\end{figure}
In this subsection, we  verify the conclusions in Section III.A through Monte Carlo simulations.

Firstly, the positions of $1800$ user data are randomly generated in the AoI, and then the corresponding RSS vectors are measured. For each user data, the prediction error of the MLP model and the minimum distance between the user position to fingerprint positions are stored as a pair. Fig.~\ref{verification}\subref{mde} is a statistical version of the $1800$ data pairs.  As  seen, the  average prediction error is in proportion to the minimum distance.

Secondly, we generate $1000$ fingerprint databases randomly. For each fingerprint database, the area of convex hull and the MAE of test database are stored as a pair. Fig.~\ref{verification}\subref{areae} is a statistical version of the $1000$ data pairs. It demonstrates that the average MAE is in inverse proportion to the area of convex hull of fingerprint database.

\subsection{Prediction Performance of Proposed FedLoc-AC}
\begin{figure}[tbp]
    \setlength{\belowcaptionskip}{-0.5cm}
	\centering
	\subfloat[Heterogeneous scenario]{\label{vhete}\includegraphics[width=0.25\textwidth]{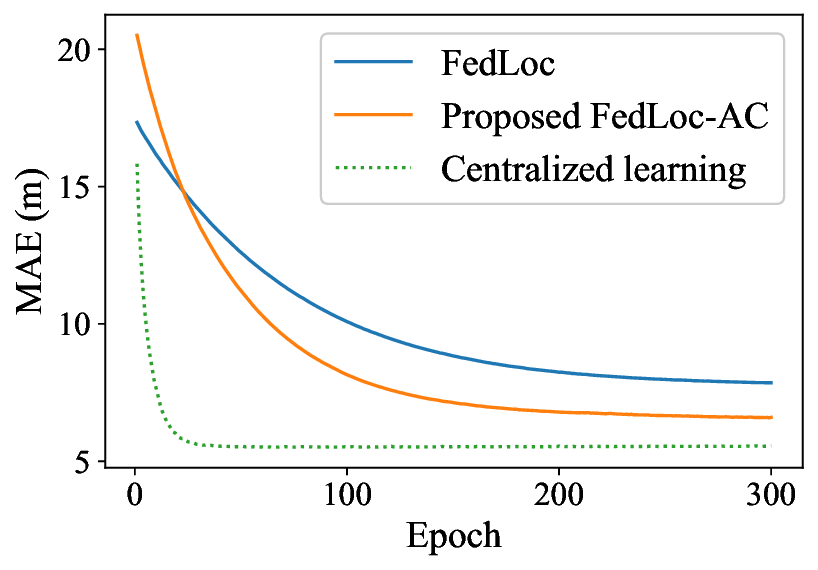}}
	\subfloat[Homogeneous scenario]{\label{nhete}\includegraphics[width=0.25\textwidth]{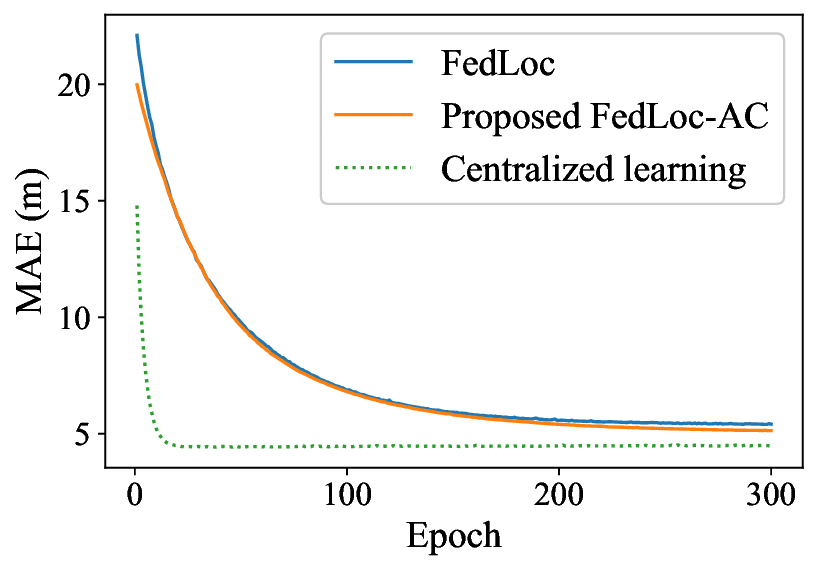}}
	\caption{Testing MAE of focused methods versus global epochs in designed scenarios.}
    \label{generatedata}
\end{figure}

Fig.~\ref{generatedata} shows the testing MAE of focused approaches versus global  epoches under the designed scenarios. In the heterogeneous scenario,    the FedLoc-AC (the proposed algorithm) surpasses the FedLoc obviously in terms of the prediction accuracy with similar convergence rate. At the final round, the FedLoc-AC  can achieve a $20\%$ performance gain compared with FedLoc. This improvement is reasonable. The fingerprint database of  clients with stronger moving abilities are more representative to reflect the environment of the AoI. The proposed FedLoc-AC distributes larger weights to the stronger clients in the model aggregation while FedLoc averages the aggregating weights. In the homogeneous scenario, the proposed FedLoc-AC has  almost the same performance with the FedLoc since the aggregating weights in FedLoc-AC are  near average. Through converging under more rounds than the centralized learning, the FedLoc-AC can keep raw data locally.

\subsection{Real-world 3D cases}
\begin{figure}
  \centering
  \includegraphics[width=0.45\textwidth]{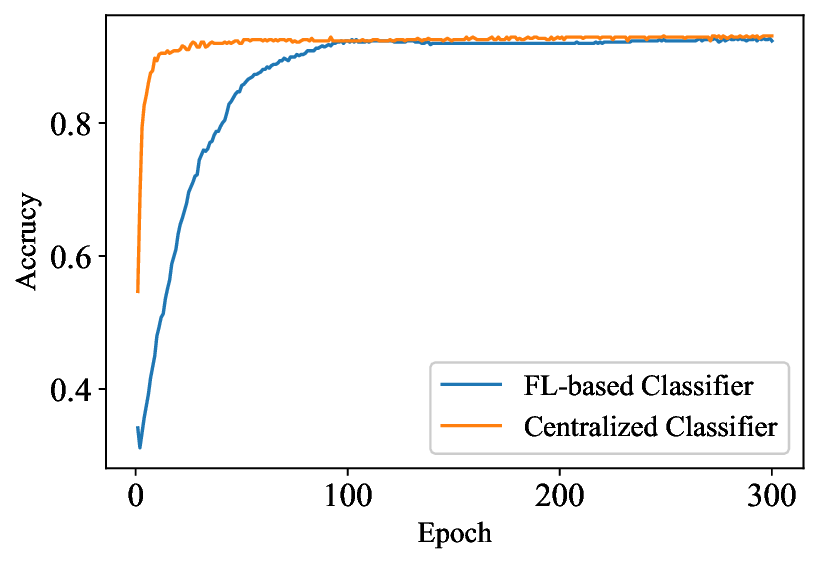}\\
  \caption{Test Accuracy of Floor Classifiers}\label{Floorsimu}
\end{figure}

\begin{figure}[tbp]
    \setlength{\belowcaptionskip}{-0.5cm}
	\centering
	\subfloat[Heterogeneous scenario]{\label{vhete}\includegraphics[width=0.25\textwidth]{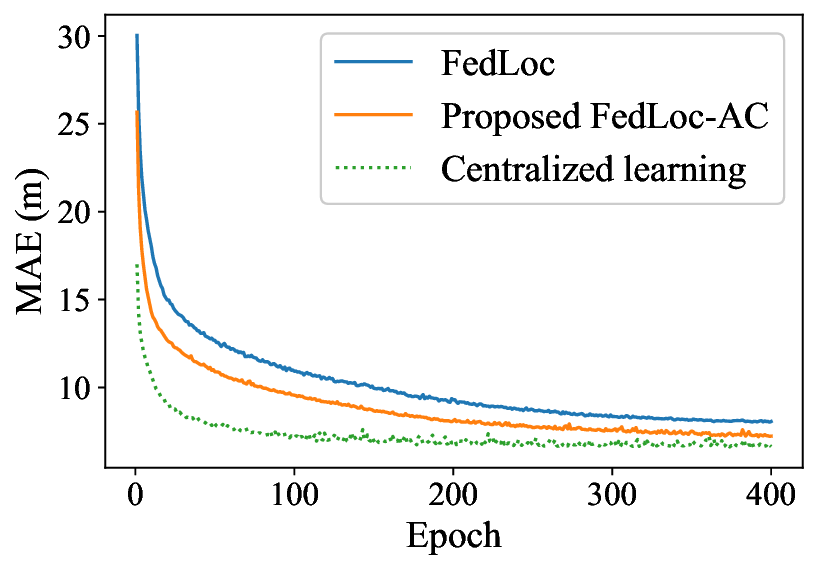}}
	\subfloat[Homogeneous scenario]{\label{nhete}\includegraphics[width=0.25\textwidth]{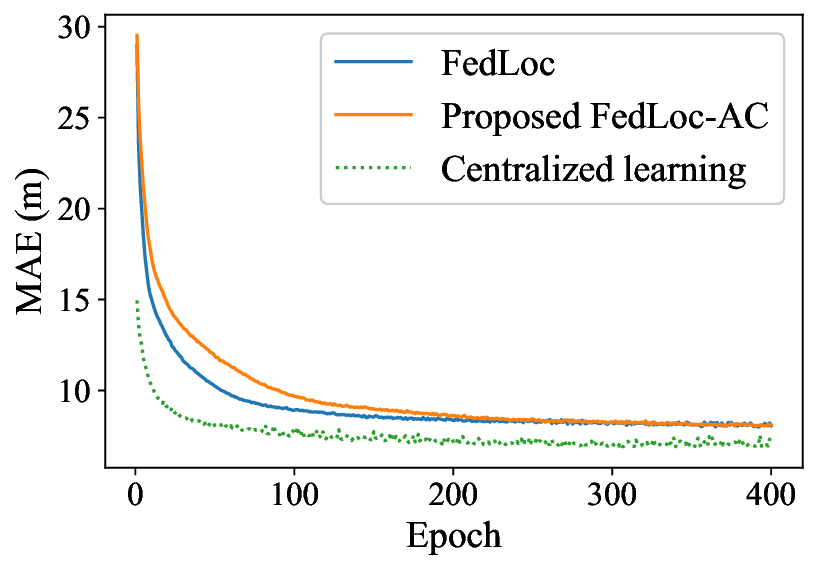}}
	\caption{Testing MAE of focused methods versus global epochs with UJIIndoorLoc database.}
    \label{UJLLocdate}
\end{figure}
The real-word dataset named UJIIndoorLoc \cite{7275492} is adopted to evaluate the proposed method and its extension version.  The experimental data is sampled in the four-storey building with ''BUILDINGID'' equal to 0  in the UJIIndoorLoc database. We select the the training subset for model training and the validation subset for test.

The client amount in the building is set to be 20. For floor classifier, we select MLP network as the classifier, consisting of a single hidden layer with 1024 hidden units, where ReLU units are selected as active function. Softmax units are selected as active function in the output layer.  The learning is achieved by  SGD with $\eta=0.000001$ and $E=20$. For floor-specific localization, the MLP is trained by  SGD consists of two hidden layers with $128\times128$  ReLU units. We set $E=40$, and $ \eta=0.00001$.

Fig.~\ref{Floorsimu} shows the testing accuracy of FL-based classifier and centralized classifier versus global epoches. As seen, the test accuracy of FL-based classifier rivals the centralized benchmark.

When the floor is predicted by the FL-based classifier, to further  predict the specific position on the floor is just the 2D case we focused on. Fig.~\ref{UJLLocdate} shows the test error of focused approaches versus global epoches on the floor with id equal to 1 in this building$\footnote{Also, we consider both  homogeneous and  heterogeneous scenarios.  In the homogeneous scenario, the training database are random distributed to 8 clients. In the heterogeneous scenario,  a half of  clients act as straggles, whose local training databases have limited cover area in the AoI.}$. This result demonstrates  the prediction superiority of the proposed  FedLoc, consisting with  Fig.~\ref{generatedata}.

\subsection{extension to 3D case}

\section{Conclusion}
In this paper, we have focused on FL-based localization that trains a global model in a cooperative and distributed manner without exposing the raw data of clients. Considering the practical database heterogeneity,  a novel  FedLoc-AC algorithm has been considered, which aggregates the  client model according to the proposed heterogeneous characteristic, i.e., the area of convex hull.  Experimental results have verified the effectiveness of the heterogeneous characteristic and confirm the prediction superiority of FedLoc-AC, compared to the existing FedLoc. Improving the FedLoc-AC by finding more effective aggregating  weights deserves further research.

\section*{Appendix:~Convergence bound of FedLoc-AC}
For theoretical analysis, the assumptions of the local loss function are listed as follows.
\begin{assumption}
We assume the following for the $i$-th client:
\begin{enumerate}
  \item $\mathcal{L}_{i}(\mathbf{w})$ is convex.
  \item $\mathcal{L}_{i}(\mathbf{w})$ is $\rho$-Lipschitz, i.e., $ \|\mathcal{L}_{i}(\mathbf{w})-\mathcal{L}_{i}(\mathbf{w}^{'}) \|\leq\rho\|\mathbf{w}-\mathbf{w}^{'}\|$  for any $\mathbf{w}, \mathbf{w}^{'}$.
  \item $\mathcal{L}_{i}(\mathbf{w})$ is $\beta$-smooth, i.e.,  $ \|\nabla \mathcal{L}_{i}(\mathbf{w})- \nabla\mathcal{L}_{i}(\mathbf{w}^{'}) \|\leq\beta \|\mathbf{w}-\mathbf{w}^{'}\|$ for any $\mathbf{w}, \mathbf{w}^{'}$.
\end{enumerate}
\end{assumption}

We also define the following metric to capture the divergence between the gradient of a local loss function, defined in (\ref{localobj}), and the gradient of the global loss function, defined in (\ref{obj}).
\begin{definition}
For any $i$ and $\mathbf{w}$, we define $\delta_{i}$ as an upper bound of  $\|\nabla \mathcal{L}_{i}(\mathbf{w})- \nabla\mathcal{L}(\mathbf{w})\|$, i.e.,
\begin{align}
\|\nabla \mathcal{L}_{i}(\mathbf{w})- \nabla\mathcal{L}(\mathbf{w})\|\leq \delta_{i}.
\end{align}
We also define $\delta\triangleq \frac{S_{C_{i}}\delta_{i}}{\sum_{i=1}^{N} S_{C_{i}}}$.
\end{definition}

The difference between the Definition 1 and the counterpart in \cite{8664630} are the definition of $\delta$, in where the proposed aggregation with heterogeneous databases are considered.

Following \cite{8664630}, when $\eta<\frac{1}{\beta}$, we have
\begin{align}
L(\mathbf{w}^{T})-L(\mathbf{w}^{\ast})\leq \frac{1}{2\eta \varphi T}+\sqrt{\frac{1}{4 \eta^2 \varphi^2 T^2}+\frac{\rho h(E) }{\eta \varphi E}}+\rho h(E),
\end{align}
where  $\varphi \triangleq \omega(1-\frac{\beta\eta}{2})$ , $\omega\triangleq\min_{t} \frac{1}{\|\mathbf{w}^(t-1)-\mathbf{w}^{\star}\|}$ and $h(x)\triangleq \frac{\delta}{\beta}((\eta \beta+1)^{x}-1)-\eta\delta x$ for any $x=0,1,2,...$.

\ifCLASSOPTIONcaptionsoff
  \newpage
\fi
\bibliographystyle{IEEEtran}
\bibliography{IEEEfull,cite}

\end{document}